\def\a{\alpha}\def\b{\beta}\def\d{\delta}
\def\f{\phi}
\def\k{\kappa}\def\l{\lambda}\def\m{\mu}\def\n{\nu}\def\r{\rho}\def\s{\sigma}
\def\y{\eta}

\def\D{\Delta}

\def\inf{\infty}\def\id{\equiv}\def\mo{{-1}}\def\ha{{1\over 2}}

\def\({\left(}\def\){\right)}\def\[{\left[}\def\]{\right]}
\def\lra{\leftrightarrow}

\def\diag{{\rm diag}}\def\tr{\triangleright}
\def\ot{\otimes}

\def\mn{{\mu\nu}}\def\ij{{ij}}

\def\st{spacetime }

\def\tran{transformations }\def\coo{coordinates }

\def\poi{Poincar\'e }

\def\ie{i.e.\ }

\def\cor{commutation relations }

\def\bc{boundary conditions }

\def\nc{noncommutative }\def\kp{$\k$-\poi }\def\km{$\k$-Minkowski }

\def\section#1{\bigskip\noindent{\bf#1}\smallskip}
\def\subsect#1{\bigskip\noindent{\it#1}\smallskip}
\def\subsection#1{\smallskip\noindent{\it#1}\smallskip}
\def\nota{\footnote{$^\dagger$}}

\def\PL#1{Phys.\ Lett.\ {\bf#1}}\def\CMP#1{Commun.\ Math.\ Phys.\ {\bf#1}}

\def\PR#1{Phys.\ Rev.\ {\bf#1}}

\def\JoP#1{J.\ Phys.\ {\bf#1}} \def\IJMP#1{Int.\ J. Mod.\ Phys.\ {\bf #1}}

\def\JHEP#1{JHEP\ {\bf#1}}\def\EPJ#1{Eur.\ Phys.\ J.\ {\bf#1}}

\def\hep#1{{\tt hep-th/#1}}

\def\ref#1{\medskip\everypar={\hangindent 2\parindent}#1}
\def\beginref{\begingroup
\bigskip
\centerline{\bf References}
\nobreak\noindent}
\def\endref{\par\endgroup}

\def\bif{}

\def\cD{{\cal D}}\def\hX{{\hat X}}\def\ku{{1\over\k}}
\def\hx{\hat x}\def\cF{{\cal F}}\def\es{extended Snyder }\def\kd{$\k$-deformed }

{\nopagenumbers
\line{}
\vskip20pt
\centerline{\bf Associative realizations of $\k$-deformed extended Snyder model}
\vskip40pt
\centerline{{\bf S. Meljanac}\nota{e-mail:\ meljanac@irb.hr}}
\vskip5pt
\centerline {Rudjer Bo\v skovi\'c Institute, Theoretical Physics Division}
\centerline{Bijeni\v cka c. 54, 10002 Zagreb, Croatia}
\vskip10pt
\centerline{and}
\vskip5pt
\centerline{{\bf S. Mignemi}\nota{e-mail:\ smignemi@unica.it}}
\vskip5pt
\centerline {Dipartimento di Matematica, Universit\`a di Cagliari}
\centerline{via Ospedale 72, 09124 Cagliari, Italy}
\smallskip
\centerline{and INFN, Sezione di Cagliari}

\vskip40pt
{\noindent\centerline{\bf Abstract}}
\vskip5pt
Usually, the realizations of the noncommutative Snyder model lead to a nonassociative star product.
However, it has been shown that this problem can be avoided by adding to the spacetime \coo new tensorial
degrees of freedom.
The model so obtained, called \es model, can be subject to a $\k$-deformation, giving rise to a unification of
the Snyder and the \kp algebras in the formalism of extended spacetime.

In this paper we review this construction and consider the generic realizations of the \kd\es model,
calculating the associated star product, coproduct and twist in a perturbative setting.
We also introduce a representation of the Lorentz algebra in the extended space and speculate on  possible
interpretations of the tensorial degrees of freedom.

\vfil\eject}
\section{1. Introduction}
Noncommutative geometries are an attempt to describe the structure of spacetime at scales of the order of the Planck length
by postulating a noncommutative structure of the operators associated to the position of a point in spacetime [1,2].
A particularly successful approach to noncommutative geometry is that based on the formalism of Hopf algebras [3], that permits to
describe the relativistic symmetries of the quantum spacetime, starting from the algebra of the position operators.
This formalism has been extensively used in particular in the investigation of the Moyal [4] and the \kp models [5].

However, not all noncommutative geometries fit easily in this framework. For example, the treatment of the Snyder model [6]
presents some difficulties, since the algebra of the position operators does not close in this case, because the commutators
are given by Lorentz generators: hence also the Lorentz generators should be included in the spacetime algebra from which the Hopf
structure is derived.
It is nevertheless possible to construct a sort of Hopf algebra even excluding the Lorentz generators, but in
this way the resulting Hopf structure does not satisfy all the standard axioms, in particular it is not coassociative [6,7]
and the related star product is not associative.

For this reason in [8], following a proposal of [7], a Hopf algebra was constructed adding to the position operators $\hx_i$
of $N$-dimensional spacetime
a set of antisymmetric rank-2 tensors $\hx_\ij$ as primary generators, in such a way that the algebra generated by the $\hx_i$ and
$\hx_\ij$ were isomorphic to the $N$-dimensional Snyder algebra. It was also shown that this algebra can be written more compactly
as an $so(1,N)$ algebra with generators $\hx_\mn$, by identifying the $\hx_i$ with the $\hx_{iN}$ of $so(1,N)$.\footnote{$^1$}{In
this paper, Greek indices run from 0 to $N$, Latin indices from 0 to $N-1$.}
To distinguish this structure from the usual realization of the Snyder model leading to nonassociativity, it was called
extended Snyder spacetime.
General realizations of the ensuing Hopf algebra in terms of an extended Heisenberg algebra were investigated
and the related coproduct and twist were calculated.

Later in [9], using a formalism developed in [10], it was shown that this construction can be extended to include also a
$\k$-deformation, by passing from an algebra $so(1,N)$ to an algebra $so(1,N;g)$ with generators $\hX_\mn$, in which the metric
$\y_\mn$ is changed into a nondiagonal metric $g_\mn$. In that paper,
only the Weyl realization of the Hopf algebra was considered and the coproduct and twist were calculated only in that case.
In particular, the determination of the twist is important also in view of applications, as for example the determination of
dispersion relations [11]. In this context, the Heisenberg double construction was investigated in [12].

The aim of this paper is to review these models and extend the investigation of the Hopf algebra of the $\k$-deformed \es spacetime
to generic realizations.
We shall assume that the general realization of the $\k$-deformed algebra can be obtained from that of the \es algebra by
changing the old variables to new ones defined by a transformation depending on the metric $g_\mn$ defined above. In this way one
can easily calculate the relevant quantities.
We recall that a unification of Snyder and \kp models was previously considered in the context of nonassociative realizations [13].

The paper is organized as follows: in sect.~2 we review the construction of the \kd \es spacetime [9] and its relation with the
\es spacetime [8]
and define the action of the Lorentz algebra on it. In sect.~3 we discuss perturbatively, up to second order in an expansion
parameter $\l$, the generic realizations of the \es algebra in terms of an extended Heisenberg algebra
generated by the operators $x_i$ and $x_\ij$ and their canonically conjugated momenta.
In sect.~4  we construct realizations of $\hX_i$ and $\hX_\ij$, the star product and the coproduct of the momenta
for a generic realization of the \kd\es model.  In sect.~5 we obtain the twist
up to first order in $\l$, while in sect.~6 the coproduct of the generators of the Lorentz \tran is computed.
In sect.~7 we draw some conclusions, {\bif while in the Appendix we recall the construction of the twist in the Hopf algebroid approach.}

\section{2. \km Snyder spacetime and $so(1,N;g)$ algebra}
\subsection{2.A. The algebras $so(1,N;g)$ and $so(1,N)$.}
In this section we review the construction of the $\k$-Minkowski-Snyder spacetime introduced in [9].
We define the algebra $so(1,N;g)$ with dimensionless generators $\hX_\mn$ as
$$[\hX_\mn,\hX_{\r\s}]=i\l(g_{\m\r}\hX_{\n\s}-g_{\m\s}\hX_{\n\r}-g_{\n\r}\hX_{\m\s}+g_{\n\s}\hX_{\m\r}),\eqno(2.1)$$
where $\l$ is a real parameter and $g_\mn$ is a symmetric matrix that plays the role of a metric.
{\bif The parameter $\l$ is dimensionless in natural units and interpolates between an abelian algebra and a $SO(1,N)$ algebra;
its actual value is arbitrary, depending on the units chosen, so that we can assume $\l\ll1$ and use it for series expansions.}

In the following, we shall be interested in metrics of the form
$$g_\mn=\left(\matrix{-1&0&\dots&0&g_0\cr
0&1&\dots&0&g_1\cr&&\dots\cr0&0&\dots&1&g_{N-1}&\cr g_0&g_1&\dots&g_{N-1}&g_N\cr}\right),\eqno(2.2)$$
with $g^T=g$ and det $g=-g_0^2+\sum_{i=1}^{N-1}g_i^2-g_N\ne0$.

If we introduce \nc\coo $\hX_i=\ku\hX_{iN}$, where $\k$ is a parameter with dimension of mass, and define $a_i=\ku g_i$,
$\b={1\over\k^2}g_N$, the relations (2.1) split into
$$\eqalignno{&[\hX_i,\hX_j]=i\l(a_i\hX_j-a_j\hX_i+\b\hX_{ij}),&\cr
&[\hX_{ij},\hX_k]=i\l(\y_{ik}\hX_j-\y_{jk}\hX_i-a_i\hX_{jk}+a_j\hX_{ik}),&\cr
&[\hX_{ij},\hX_{kl}]=i\l(\y_{ik}\hX_{jl}-\y_{il}\hX_{jk}-\y_{jk}\hX_{il}+\y_{jl}\hX_{ik}).&(2.3)}$$
In this way we obtain a unification of extended Snyder and \km spacetimes using the formalism of extended \coo $\hX_\mn$.
The $\hX_i$ are interpreted as spacetime coordinates, while the $\hX_\ij$ are new tensorial degrees of freedom, whose physical
interpretation we leave open.
The algebra (2.1) generated by the $\hX_\mn$ reduces to the \kp algebra if $g_N=0$, and to the Snyder algebra
if $g_0=\dots=g_{N-1}=0$.

In the latter case, the algebra (2.1) reduces to the standard $so(1,N)$ algebra with generators $\hx_\mn$, considered in [8]
and defined by
$$[\hx_\mn,\hx_{\r\s}]=i\l(\y_{\m\r}\hx_{\n\s}-\y_{\m\s}\hx_{\n\r}-\y_{\n\r}\hx_{\m\s}+\y_{\n\s}\hx_{\m\r}).\eqno(2.4)$$
where $\y_\mn=\diag(-1,1,\dots,1)$, and we have set $g_N=1$.

Clearly, if one defines, in analogy with the previous case, $\hx_i=\ku\hx_{iN}$, {\bif and identifies ${1\over\k^2}$ with $\b$
(\ie sets $g_N=1)$}, the \cor (2.4) take the form of the \es algebra since
$$\eqalignno{&[\hx_i,\hx_j]=i{\l\over\k^2}\hx_{ij},&\cr
&[\hx_{ij},\hx_k]=i\l(\y_{ik}\hx_j-\y_{jk}\hx_i),&\cr
&[\hx_{ij},\hx_{kl}]=i\l(\y_{ik}\hx_{jl}-\y_{il}\hx_{jk}-\y_{jk}\hx_{il}+\y_{jl}\hx_{ik}).&(2.5)}$$

One can write the relation between the $\hX_\mn$ and $\hx_\mn$ and between the metrics $g_\mn$ and $\y_\mn$
by means of a matrix $O$ such that
$$\hX_\mn=(O\hx O^T)_\mn,\qquad g_\mn=(O\y O^T)_\mn,\eqno(2.6)$$
with
$$O_\mn=\left(\matrix{1&0&\dots&0&0\cr0&1&\dots&0&0\cr&&\dots\cr0&0&\dots&1&0\cr
-g_0&g_1&\dots&g_{N-1}&\r\cr}\right),\eqno(2.7)$$
where $\r=\sqrt{g_N-g_kg_k}=\sqrt{-\det g}$. The matrix $O$ is defined up to $SO(1,N)$ transformations.

Then
$$\hX_i=\r\,\hx_i+a_j\,\hx_{ij},\qquad\hX_{ij}=\hx_{ij},\eqno(2.8)$$
with inverse
$$\hx_i={1\over\r}\,(\hX_i-a_j\,\hX_{ij}),\qquad\hx_{ij}=\hX_{ij}.\eqno(2.9)$$

An important limit is obtained for $\l=0$: {\bif in this limit, one is left with an abelian algebra, \ie a commutative extended spacetime,
from which it is possible to construct an extended Heisenberg algebra}.
In fact, calling $X_\mn$ and $x_\mn$ the commuting operators corresponding to $\hX_\mn$ and $\hx_\mn$ in this limit,
we can  define their canonically conjugate momenta $P^\mn$ and $p^\mn$, respectively. These are also related by the matrix $O$ as
$$P^\mn=(O^\ddagger\,p\,O^\mo)^\mn,\eqno(2.10)$$
where
$$O^\ddagger\id(O^\mo)^T={1\over\r}\left(\matrix{\r&0&\dots&0&g_0\cr0&\r&\dots&0&-g_1\cr&&\dots\cr0&0&\dots&\r&-g_{N-1}\cr
0&0&\dots&0&1\cr}\right),\eqno(2.11)$$
or, more explicitly\footnote{$^2$}{We adopt the convention that Latin indices are lowered and raised by the flat metric, thus in
the following we shall always write them in lower position.} defining $P_i=\k\,P^{iN}$, and $p_i=\k\,p^{iN}$,
$$P_i={1\over\r}\,p_i,\qquad P_\ij=p_\ij+{1\over\r}(a_ip_j-a_jp_i).\eqno(2.12)$$
The inverse relations are
$$p_i=\r P_i,\qquad p_\ij=P_\ij-a_iP_j+a_jP_i.\eqno(2.13)$$

Together with the respective position variables, the momenta satisfy the extended Heisenberg algebras
$$[X_\mn,X_{\r\s}]=[P^\mn,P^{\r\s}]=0,\qquad[X_\mn,P^{\r\s}]=i(\d_\m^{\ \r}\d_\n^{\ \s}-\d_\m^{\ \s}\d_\n^{\ \r}),\eqno(2.14)$$
and
$$[x_\mn,x_{\r\s}]=[p^\mn,p^{\r\s}]=0,\qquad[x_\mn,p^{\r\s}]=i(\d_\m^{\ \r}\d_\n^{\ \s}-\d_\m^{\ \s}\d_\n^{\ \r}),\eqno(2.15)$$
respectively. Moreover, analogs of the relations (2.6) and (2.8) are valid for $X_\mn$ and $x_\mn$.
$$X_\mn=(OxO^T)_\mn,\qquad X_i=\r\,x_i+a_j\,x_{ij},\qquad X_{ij}=x_{ij},\eqno(2.16)$$
with $X_i=\ku X_{iN}$ and $x_i=\ku x_{iN}$, and inverse
$$x_i={1\over\r}\,(X_i-a_j\,X_{ij}),\qquad x_{ij}=X_{ij}.\eqno(2.17)$$

{\bif The relevance of this formalism is that it permits to pass from the standard results relative to the \es model to those of its \kd
extension by simply transforming the $X$, $P$ variables to the $x$, $p$ ones with (2.8), (2.12) and use the formulas valid in this case,
and then go back to the \kd variables by means of (2.9), (2.13), obtaining the corresponding deformed relations.}

\subsect{2.B. The Lorentz algebra acting on the extended Heisenberg algebra.}
One can also define an $N$-dimensional Lorentz algebra $so(1,N-1)$ acting on the \coo $x_\mn$
in such a way that the $x_i$ transform as $N$-vectors and the $x_\ij$ as antisymmetric tensors.
The generators of the Lorentz algebra are given by the operators $M_\ij$ defined as
$$M_\ij=x_ip_j-x_jp_i+x_{ik}p_{jk}-x_{jk}p_{ik},\eqno(2.18)$$
and satisfy
$$[M_\ij,M_{kl}]=i(\y_{ik}M_{jl}-\y_{il}M_{jk}-\y_{jk}M_{il}+\y_{jl}M_{ik}),\eqno(2.19)$$
together with
$$\eqalignno{&[M_\ij,x_k]=i(\y_{ik}x_j-\y_{jk}x_i),\qquad[M_\ij,x_{kl}]=i(\y_{ik}x_{jl}-\y_{jk}x_{il}-\y_{il}x_{jk}+\y_{jl}x_{ik}),\cr
&[M_\ij,p_k]=i(\y_{ik}p_j-\y_{jk}p_i),\qquad[M_\ij,p_{kl}]=i(\y_{ik}p_{jl}-\y_{jk}p_{il}-\y_{il}p_{jk}+\y_{jl}p_{ik}).&(2.20)}$$

It is also easy to check that
$$[M_\ij,\hx_k]=i(\y_{ik}\hx_j-\y_{jk}\hx_i),\qquad[M_\ij,\hx_{kl}]=i(\y_{ik}\hx_{jl}-\y_{jk}\hx_{il}-\y_{il}\hx_{jk}+\y_{jl}\hx_{ik}).\eqno(2.21)$$

\section{3. Generic realization and star product for the extended Snyder model}
\subsection{3.A. Generic realization of the extended Snyder model}
We now recall how the extended Snyder model (2.5) can be realized in terms of the extended Heisenberg algebra (2.15) [8].
A realization of the model is defined as
$$\hat x_i=x_k\f_1(\l p)_{k,i}+x_{kl}\f_2(\l p)_{kl,i},\qquad\hat x_\ij=x_k\f_3(\l p)_{k,ij}+x_{kl}\f_4(\l p)_{kl,ij},\eqno(3.1)$$
where the matrix functions $\f(\l p)$ satisfy the differential equations that can be obtained from (2.5) with \bc $\hat x_i=x_i$, $\hat x_\ij=x_\ij$ for $\l=0$.

In [8] the most general realization of these commutation relations in terms of elements of the extended Heisenberg algebra (2.15) up to second order in $\l$
was shown to depend on 14 parameters, namely
$$\eqalignno{\hat x_i&=x_i+\l\big[\b c_0x_{ik}p_k+c_1x_kp_{ik}\big]+\l^2\big[\b(c_2x_ip_k^2+c_3x_kp_kp_i+c_4x_{ik}p_{kl}p_l+c_5x_{kl}p_kp_{il})+c_6x_kp_{kl}p_{il}\big],\cr
\hat x_\ij&=x_\ij+\l\big[d_0x_{ik}p_{jk}+d_1x_ip_j-(i\lra j)\big]+\l^2\big[\b d_2x_{ik}p_kp_j+d_3x_{ik}p_{kl}p_{jl}+d_4x_{kl}p_{ik}p_{jl}+d_5x_ip_kp_{jk}\qquad&\cr
&\quad+d_6x_kp_{ik}p_j-(i\lra j)\big],&(3.2)}$$
where the coefficients of the first-order terms must satisfy
$$c_0=-\ha,\qquad d_0=\ha,\qquad c_1+d_1=1,\eqno(3.3)$$
while those of the second-order terms satisfy six further independent relations,
$$\eqalignno{&\ {c_1\over2}-2c_2+c_3=d_1,\qquad{c_1\over2}+c_4+c_5=\ha,\qquad d_3-2d_4=-{1\over4},&\cr
&c_5-d_2={1\over4},\qquad{c_1\over2}+c_6-d_6=0,\qquad{c_1\over2}-c_1d_1+c_6+d_5=0.&(3.4)}$$
Therefore, to first order in $\l$ a single free parameter, $c_1$, is left, while up to second order one has five free parameters.
Note that setting $\b=0$ in (3.2) one obtains realizations of the \poi algebra.
In particular, for $\b=0$, $c_1=c_6=0$, it follows $\hx_i=x_i$, describing ordinary Minkowski space.

\subsect{3B. The star product for the extended Snyder model}
{\bif To define the  star product for a generic realization of the kind $\hx_\a=x_\b\f_{\b\a}(p)$, where $\a$, $\b$ are arbitrary indices,
using the action $\tr$ we calculate [13]
$$e^{ik\hx}\tr1=e^{iK(k)x}.\eqno(3.5)$$
Then
$$e^{ikx}=e^{iK^\mo(k)\hx}\tr1,\eqno(3.6)$$
and the star product of two plane waves is defined as
$$e^{ikx}\star e^{iqx}=e^{iK^\mo(k)\hx}\tr e^{iqx}\id e^{i\cD(k,q)x}.\eqno(3.7)$$}

Applying this definition to the realizations (3.2) we obtain
$$e^{ik_ix_i+{i\over2} k_\ij x_\ij}\star e^{iq_kx_k+{i\over2}q_{kl}x_{kl}}=e^{i\cD_ix_i+{i\over2}\cD_\ij x_\ij},\eqno(3.8)$$
where [8]
$$\eqalignno{\cD_i(k,q)&=k_i+q_i+\l\big[-c_1k_jq_\ij+d_1k_{ij}q_j\big]+{\l^2\over2}\big[\,\b(c_0c_1+c_3)k_j^2q_i+\b(-c_0c_1+2c_2+c_3)k_ik_jq_j&\cr
&+(c_1^2-c_1d_0-c_1d_1+c_6+d_6)k_jk_{jk}q_{ik}+(c_1d_0+c_1d_1+c_6-d_5)k_{ik}k_jq_{jk}&\cr
&+(d_1^2+d_5-d_6)k_\ij k_{jk}q_k+2\b c_2k_iq_j^2+2\b c_3k_jq_jq_i+2c_6k_jq_{jk}q_{ik}+2d_5k_{ij}q_{jk}q_k&\cr
&+2d_6k_{jk}q_jq_{ik}\big]+o(\l^3),\cr
\cD_\ij(k,q)&=k_\ij+q_\ij+\l\big[-d_0k_{ik}q_{jk}+\b c_0k_iq_j-(i\lra j)\big]+{\l^2\over2}\big[\,\b(-c_0c_1+c_4-c_5)k_ik_kq_{jk}&\cr
&+(-d_0^2+d_3+2d_4)k_{ik}k_{kl}q_{jl}+\d_0^2+d_3)k_{ik}k_{jl}q_{kl}+\b(c_0d_0+c_5+d_2)k_{ik}k_kq_j&\cr
&+\b(c_0d_0+c_0d_1+c_4-d_2)k_ik_{jk}q_k+2\b d_2k_{ik}q_kq_j+2d_3k_{ik}q_{kl}q_{jl}+2d_4k_{kl}q_{ik}q_{jl}&\cr
&+2\b c_4k_iq_kq_{jk}+2\b c_5k_kq_{ik}q_j-(i\lra j)\big]+o(\l^3).&(3.9)}$$
The star product so defined is associative.

\section{4. Generic realization and star product for the $\k$-deformation}
\subsection{4.A. Realization of $\hX_i$ and $\hX_\ij$.}
In this section we extend the previous results to the $\k$-deformation of the model, employing the formalism introduced in sect.~2. For simplicity,
we shall consider only expansions up to order $\l$.

We assume that the generators $\hX_i$ and $\hX_\ij$ of the $\k$-deformed extended Snyder model in a generic realization can
be obtained in terms of the $x$ and $p$ by substituting (2.8) and (2.12) in (3.2). Taking into account the relations (3.3), at first order in $\l$ we get
$$\eqalignno{\hX_i=&\ \r\hx_i+\hx_\ij a_j=\r x_i+x_\ij a_j+{\l\over2}\,\Big[-\b\r\,x_{ij}p_j+2c_1\r\,x_jp_{ij}+(x_{ik}p_{jk}-x_{jk}p_{ik})a_j\cr
&+2(1-c_1)(x_ip_ka_k-x_kp_ia_k)\Big],\cr
\hX_\ij=&\ \hx_\ij=x_\ij+{\l\over2}\,\Big[x_{ik}p_{jk}+2(1-c_1)x_ip_j-(i\lra j)\Big].&(4.1)}$$

Using (2.9) and (2.13), the general realization can then be written in terms of $X_i$, $X_\ij$, $P_i$ and $P_\ij$,
$$\eqalignno{\hX_i=&\ X_i+{\l\over2}\,\Big[-\b\,X_{ij}P_j+2c_1(X_jP_{ij}-X_jP_ja_i)+2(1-c_1)X_iP_ja_j+X_{ik}P_{jk}a_j\cr
&+(2c_1-1)\(2X_jP_ia_j-X_{jk}P_\ij a_k+X_{ik}P_ja_ka_j+X_{jk}P_ja_ka_i\)\Big],\cr
\hX_\ij=&\ X_\ij+{\l\over2}\,\Big[X_{ik}P_{jk}+2(1-c_1)X_iP_j-X_{ik}P_ka_j+(2c_1-1)X_{ik}P_ja_k-(i\lra j)\Big].&(4.2)}$$
{\bif Note that in the limit $\k\to\inf$, \ie $a_i=\b=0$, in (4.2) one obtains realizations of the \poi algebra. In particular if also $c_1=0$,
it follows that $\hX_i=X_i$ spans the ordinary Minkowski spacetime, which decouples from the tensorial coordinates.

The previous limit implies $g_0=\dots=g_{N-1}=g_N=0$. In this case the matrices $g$ and $O$ are singular and do not have an inverse
 and (2.8), (2.9), as well as (4.1) do not make sense.
If instead $g_N=0$, $g_k^2=0$, but $g_i\ne0$, it follows that $\r=0$, $\det O=\det g=0$ and $\hX_i=a_j\hx_\ij$, corresponding to a realization
of lightlike \km spacetime [14].
Finally, if $g_N=0$, $g_i\ne0$, $g_k^2<0$ it follows that $\r\ne0$, $\det O\ne0$, $\det g\ne0$ and
the corresponding algebra is isomorphic to the timelike \kp algebra.}

\subsect{4.B. The star product and coproduct for the $\k$-deformation}
The star product of exponentials in terms of \coo $X_i$, $X_\ij$ and momenta $K_i$, $K_\ij$ and $Q_i$, $Q_\ij$ is defined as
$$e^{iK_iX_i+{i\over2}K_\ij X_\ij}\star e^{iQ_kX_k+{i\over2}Q_{kl}X_{kl}}=e^{i\cD_i(K,Q)X_i+{i\over2}\cD_\ij(K,Q) X_\ij},\eqno(4.3)$$
Using the relations (2.16) and (2.12) it follows that
$$K_iX_i+\ha K_\ij X_\ij=k_ix_i+\ha k_\ij x_\ij,\qquad Q_iX_i+\ha Q_\ij X_\ij=q_ix_i+\ha q_\ij x_\ij,\eqno(4.4)$$
and therefore
$$e^{iK_iX_i+{i\over2}K_\ij X_\ij}\star e^{iQ_kX_k+{i\over2}Q_{kl}X_{kl}}=e^{ik_ix_i+{i\over2} k_\ij x_\ij}\star e^{iq_kx_k+{i\over2}q_{kl}x_{kl}}
=e^{i\cD_i(k,q)x_i+{i\over2}\cD_\ij(k,q)x_\ij},\eqno(4.5)$$
Using then the inverse relations (2.17), we obtain
$$\cD_i(K,Q)={1\over\r}\cD_i(k,q),\qquad \cD_\ij(K,Q)=\cD_\ij(k,q)+{1\over\r}\Big(a_i\cD_j(k,q)-a_j\cD_i(k,q)\Big).\eqno(4.6)$$
Finally, from (3.9) and the relations (2.13) for $k_i$, $k_\ij$ and $q_i$, $q_\ij$, we get
$$\eqalignno{\cD_i(K,Q)=&\ K_i+Q_i+\l\,\Big[-c_1(K_jQ_{ij}+K_jQ_ia_j)+(1-c_1)(K_{ij}Q_j+K_iP_ja_j)+(2c_1-1)K_jQ_ja_i\Big],\cr
\cD_\ij(K,Q)=&\ K_\ij+Q_\ij+{\l\over2}\,\Big[-\b K_iQ_j-(K_{ik}Q_{jk}+K_iQ_{jk}a_k+K_{jk}Q_ia_k)-(2c_1-1)(K_kQ_{jk}a_i\cr
&+K_kQ_ja_ka_i+K_{jk}Q_ka_i+K_jQ_ka_ka_i)-(i\lra j)\Big].&(4.7)}$$

The coproduct is then obtained as {\bif (see Appendix)}
$$\D P_i=\cD_i(P\ot1+1\ot P)\qquad\D P_\ij=\cD_\ij(P\ot1+1\ot P).\eqno(4.8)$$
At first order in $\l$ we have therefore
$$\eqalignno{\D P_i=&\ \D_0P_i+\l\,\Big[-c_1(P_j\ot P_{ij}+P_j\ot P_ia_j)+(1-c_1)(P_{ij}\ot P_j+P_i\ot P_ja_j)+(2c_1-1)P_j\ot P_ja_i\Big],\cr
\D P_\ij=&\ \D_0P_\ij+{\l\over2}\,\Big[-\b P_i\ot P_j-(P_{ik}\ot P_{jk}+P_i\ot P_{jk}a_k+P_{jk}\ot P_ia_k)
-(2c_1-1)(P_k\ot P_{jk}a_i\cr&+P_k\ot P_ja_ka_i+P_{jk}\ot P_ka_i+P_j\ot P_ka_ka_i)-(i\lra j)\Big].&(4.9)}$$
where
$$\D_0P_i=P_i\ot1+1\ot P_i,\qquad\D_0P_\ij=P_\ij\ot1+1\ot P_\ij.\eqno(4.10)$$
This coproduct is coassociative,
$$(\D\ot1)\D=(1\ot\D)\D.\eqno(4.11)$$

Finally, note that $\D P_i={1\over\r}\D p_i$ and  $\D P_\ij=\D\(p_\ij+{1\over\r}(a_ip_j-a_jp_i)\)$,
and after inserting (2.12) in these relations, we get the expressions (4.9) for $\D P_i$ and $\D P_\ij$.
This is consistent with the results (4.7) for $\cD_i(K,Q)$ and $\cD_\ij(K,Q)$.


\section{5. The twist}
In the Hopf algebroid approach, the twist is defined as a bilinear operator such that $\D h=\cF\D_0h\cF^\mo$ for each $h$ in the algebroid,
and it has been discussed in several papers [15,16], {\bif see Appendix A}. It is given by
$$\cF^\mo=\ :e^{{i\over2}(1\ot X_\mn)(\D-\D_0)P^\mn}\!:\eqno(5.1)$$
where $:\ :$ denotes normal ordering [16]. {\bif Note that this definition of twist differs from that of Drinfeld twist in the Hopf algebra sense,
since in the latter case the coordinates $X$ do not appear.}

In [9], the twist for the \kd\es model was calculated perturbatively for a Weyl realization.
Here we compute it for a generic realization: defining $\cF^\mo=e^F$, one gets at leading order
$$\eqalignno{F&=i(1\ot X_i)(\D-\D_0)P_i+{i\over2}(1\ot X_\ij)(\D-\D_0)P_\ij\cr &=iP_i\ot(\hX_i-X_i)+{i\over2}P_\ij\ot(\hX_\ij-X_\ij),&(5.2)}$$
which in the Weyl realization yields [9]
$$F={i\l\over2}\Big[P_i\ot\Big(-\b X_{ij}P_j+X_jP_{ij}-X_jP_ja_i+X_i P_ja_j+X_\ij P_{kj}a_k\Big)+P_\ij\ot\Big(X_{ik}P_{jk}+X_iP_j-X_{ik}P_ka_j\Big)\Big].
\eqno(5.3)$$

In the general case, using either (4.2) or (4.9), we obtain
$$\eqalignno{F ={i\l\over2}\,\Big[&P_i\ot\Big(-\b\,X_{ij}P_j+2c_1(X_jP_{ij}-X_jP_ja_i)+2(1-c_1)X_iP_ja_j+X_{ik}P_{jk}a_j\cr
&+(2c_1-1)\(2X_jP_ia_j-X_{jk}P_\ij a_k+X_{ik}P_ja_ka_j+X_{ik}P_ja_ka_j\)\Big)\cr
&+P_\ij\ot\Big(X_{ik}P_{jk}+2(1-c_1)X_iP_j-X_{ik}P_ka_j+(2c_1-1)X_{ik}P_ja_k)\Big)\Big].&(5.4)}$$
It is easy to check that
$$\D P^i=\cF\D_0P^i\cF^\mo,\qquad\D P^\ij=\cF\D_0P^\ij\cF^\mo.\eqno(5.5)$$
It also holds
$$\hX_i=m\cF^\mo(\tr\ot1)(X_i\ot1),\qquad\hX_\ij=m\cF^\mo(\tr\ot1)(X_\ij\ot1).\eqno(5.6)$$

\section{6. The Lorentz algebra}
The Lorentz algebra $so(1,N-1)$ with generators $M_\ij$ is defined in (2.19).
From (2.8) and (2.12) one obtains
$$\eqalignno{[M_\ij,X_k]&=i(\y_{ik}X_j-\y_{jk}X_i+a_jX_{ik}-a_iX_{jk}),\cr
[M_\ij,X_{kl}]&=i(\y_{ik}X_{jl}-\y_{jk}X_{il}-\y_{il}X_{jk}+\y_{jl}X_{ik}),&(6.1)}$$
where now
$$M_\ij=X_iP_j+X_{ik}P_{jk}-a_jX_{ik}P_k-(i\lra j).\eqno(6.2)$$
The definition of $M_\ij$ is obtained by the previous recipe of substituting (2.17), (2.13) in (2.18)
In the same way we obtain
$$\eqalignno{[M_\ij,P_k]&=i(\y_{ik}P_j-\y_{jk}P_i),\cr
[M_\ij,P_{kl}]&=i\Big[\y_{ik}(P_{jl}-a_jP_l)-\y_{il}(P_{jk}-a_jP_k)-(i\lra j)\Big].&(6.3)}$$

It is easy to see that the relations (6.1) also hold for $\hX_i$, $\hX_\ij$ in a generic realization,
$$\eqalignno{[M_\ij,\hX_k]&=i(\y_{ik}\hX_j-\y_{jk}\hX_i+a_j\hX_{ik}-a_i\hX_{jk}),\cr
[M_\ij,\hX_{kl}]&=i(\y_{ik}\hX_{jl}-\y_{jk}\hX_{il}-\y_{il}\hX_{jk}+\y_{jl}\hX_{ik}).&(6.4)}$$
The \cor of the $\hX_i$, $\hX_\ij$ and $M_\ij$ satisfy all the relevant Jacobi identities and generate a Lie algebra.
We have defined the action of these operators on the unity as
$$\hX_i\tr1=X_i,\qquad\hX_\ij\tr1=X_\ij,\qquad M_\ij\tr1=0,\eqno(6.5)$$

Interesting special cases are $g_N=0$, that corresponds to \km spacetime with extra tensorial \coo and momenta,
and $g_i=0$, that corresponds to the extended Snyder model.

Using the twist (5.4) we can calculate
$$\D M_\ij=\D_0M_\ij+o(\l^2),\eqno(6.6)$$
where $\D_0M_\ij=1\ot M_\ij+M_\ij\ot1$, since order-$\l$ corrections vanish.

We can improve this result, noting that if $a_i=0$, the twist (5.1) can be written as
$$\cF=\exp\(\sum_{k,l=1}^\inf\l^{k+l-1}f_{k,l}\),\eqno(6.7)$$
where $f_{k,l}\sim p^{k}\ot xp^l$, with all indices contracted. Then, $[f_{k,l},M_\ij\ot1+1\ot M_\ij]=0$ for any $k$, $l$ and hence
$\D M_\ij=\D_0M_\ij$. Moreover, our construction implies that also for $a_i\ne0$, $\D M_\ij=\D_0M_\ij$.

\section{7. Conclusions}
In this paper we have discussed the $\k$-deformation of the \es model, by introducing a deformation of the
flat metric in the definition of the \es algebra [9]. This formalism permits to have an associative star product
and a coassociative coproduct, contrary to the standard implementations of the \kd Snyder model [13].
Using this formalism, we were able to calculate in a straightforward way several properties of the associated Hopf algebra.
We also introduced a suitable definition of the Lorentz algebra acting
on the extended space.

Before concluding, we explicitly relate the parameters of our model to the standard ones in \km and Snyder:
in the present paper we have introduced the dimensionless parameters $\l$, and $g_i$, $g_N$ coming from the metric
$g_\mn$, together with a mass parameter $\k$.
The timelike \km \st is described by
$$[\hx_0,\hx_i]=-{i\over\tilde\k}\hx_i,\qquad[\hx_i,\hx_j]=0,\qquad i=1,\dots,N-1,\eqno(7.1)$$
while the original Snyder model is described by
$$[\hx_i,\hx_j]={i\over m^2}M_\ij,\eqno(7.2)$$
where $M_\ij$ are Lorentz generators and $m$ is a constant of the order of the Planck mass.

We assume that the matrix $O$ is real. Consequently, $\r^2=g_N-g_kg_k>0$.
Let us define
$$a_i={g_i\over\k}={|g|u_i\over\k},\eqno(7.3)$$
where $|u_i|\le1$, $|u_ku_k|=1$ and $|g|=\sqrt{g_kg_k}\ne0$.
Then, comparing (2.3) with (7.1) we obtain
$${1\over\tilde\k}={\l|g|\over\k},\eqno(7.4)$$
while comparing (2.3) with (7.2),
$${1\over m^2}={\l g_N\over\k^2}.\eqno(7.5)$$
Hence,
$${\tilde\k\over m}=\sqrt{g_N\over\l|g|^2}.\eqno(7.6)$$
If $m$ is of the order of the Planck mass, and $\tilde\k<m$, $g_N<\l g_k^2$.
Note that in the limit $|g|=0$ it follows $\tilde\k=0$ and we are left with the \es model.
If instead $g_N=0$ with $a^2<0$ the corresponding algebra is isomorphic to the timelike \kp algebra.

From a physical point of view, the major problem of our formalism is the interpretation of the tensorial degrees of freedom corresponding to the
 coordinates $\hX_\ij$. When $N=4$,
one may interpret them as parametrizing six extra dimensions. In the limit $\k\to\inf$, the four-dimensional spacetime and the internal
space are completely independent. However, at Planck-scale distances or energies these two spaces mix, giving rise to possible physical effects.

This is even more evident if one assigns the $\hX_\ij$ the dimension of a length, like the $\hX_i$, in contrast with the previous
discussion where we have implicitly assumed that the \coo $\hX_{ij}$ have dimension 0 in natural units, as in the original Snyder model.
However, no compelling reason constrains the choice of the physical dimension of the tensorial degrees of freedom, which depends on their physical
interpretation. With this choice, $\l$ becomes dimensional and we may identify it with the Planck length.
One may then interpret the $X_{ij}$ as position coordinates, that parametrize a disconnected universe, that interacts
with ours very weakly only at the Planck scale. However, its gravitational effects are still effective and may contribute to the dark matter content
of our universe.

At the present level of development of the theory the previous considerations are only conjectural.
Of course, a more elaborate model, taking into account also the dynamics would be necessary to discuss quantitatively this hypothesis.

\section{Acknowledgements}
We wish to thank J. Lukierski and A. Pachol for interesting discussions.
\bigskip
\section{Appendix: Twist in the Hopf algebroid approach}
The twist in the Hopf algebroid approach [16] is constructed so that the following relation holds:
$$e^{ikx}\star e^{iqx}=m\cF^\mo(\tr\ot\tr)(e^{ikx}\ot e^{iqx})=e^{i\cD(k,q)x}.\eqno(A.1)$$
The result for the twist is, using normal ordering,
$$\cF^\mo=:e^{(1\ot x)(\D-\D_0)p}:,\eqno(A.2)$$
with action
$$\D p(\tr\ot\tr)(e^{ikx}\ot e^{iqx})=\cD(k,q)(e^{ikx}\ot e^{iqx}).\eqno(A.3)$$

Hence,
$$\D p_\m=\cD_\m(p\ot1,1\ot p),\eqno(A.4)$$
where $\D p_\m$ is the coproduct of the momentum $p_\m$ since
$$\D p_\m=\cF\D_0p_\m\cF^\mo,\qquad{\rm with}\quad\D_0p_\m=(p_\m\ot1,1\ot p_\m).\eqno(A.5)$$
Since the star product is associative, the coproduct so defined is  coassociative.
\bigskip
\beginref
\ref[1] H.S. Snyder, \PR{71}, 38 (1947).
\ref[2] S. Doplicher, K. Fredenhagen and J. E. Roberts, \PL{B331}, 39 (1994); S. Doplicher, K. Fredenhagen and J. E. Roberts, \CMP{172}, 187 (1995).
\ref[3] S. Majid, {\it Foundations of quantum group theory}, Cambridge Un. Press 1995.
\ref[4] V.P. Nair and A.P. Polychronakos, \PL{B505}, 267 (2001); L. Mezincescu, \hep{0007046}.
\ref[5] J. Lukierski, H. Ruegg, A. Novicki and V.N. Tolstoi, \PL{B264}, 331 (1991); J. Lukierski and  H. Ruegg, \PL{B329}, 189 (1994).
 S. Zakrewski, \JoP{A27}, 2075 (1994); S. Majid and  H. Ruegg, \PL{B334}, 348 (1994).
\ref[6] M.V. Battisti and S. Meljanac, \PR{D82}, 024028 (2010); M.V. Battisti and S. Meljanac, \PR{D79}, 067505 (2009).
\ref[7] F. Girelli and E. Livine, \JHEP{1103}, 132 (2011).
\ref[8] S. Meljanac and S. Mignemi, \PR{D102}, 126011 (2020).
\ref[9] S. Meljanac and S. Mignemi, \PL{B814}, 136117 (2021).
\ref[10] A. Borowiec and A. Pachol, \EPJ{C74}, 2812  (2014).
\ref[11] D. Meljanac, S. Meljanac, S. Mignemi and R. \v Strajn, \PR{D99}, 126012 (2019);
D. Meljanac, S. Meljanac, Z. \v Skoda and R. \v Strajn, \IJMP{A35}, 2050034 (2020).
\ref[12] S. Meljanac and A. Pachol, Symmetry {\bf 13}, 1055 (2021).
\ref[13] S. Meljanac, D. Meljanac, A. Samsarov and M. Stoji\'c, \PR{D83}, 065009 (2011).
\ref[14] T. Juri\'c, S. Meljanac, D. Pikuti\'c and R. \v Strajn, JHEP {\bf 1507}, 055 (2015);
T. Juri\'c, S. Meljanac and A Samsarov, J.Phys.Conf.Ser. {\bf 634}, 012005 (2015).
\ref[15] T. Juri\'c, S. Meljanac and R. \v Strajn, \PL{A377}, 2472 (2013);
T. Juri\'c, D. Kovacevic and S. Meljanac, SIGMA{\bf 10}, 106 (2014);
J. Lukierski, D. Meljanac, S. Meljanac, D. Pikuti\'c, and M. Woronowicz, \PL{B777}, 1 (2018).
\ref[16] S. Meljanac, D. Meljanac, A. Pachol, D. Pikuti\'c, \JoP{A50}, 265201 (2017);
S. Meljanac, D. Meljanac, S. Mignemi and R. \v Strajn, \IJMP{A32}, 1750172 (2017).
\endref

\end